\documentclass[english,aps, superscriptaddress, floatfix, 12pt]{revtex4}
\usepackage[T1]{fontenc}
\usepackage[latin1]{inputenc}
\usepackage{graphicx}
\usepackage{amssymb}
\usepackage{epsfig}
\usepackage{amsmath}
\usepackage{psfrag}

\usepackage{babel}
\makeatother
\begin{document}

\title{Shear viscosity of the gluon plasma in the stochastic-vacuum approach}

\author{Dmitri Antonov\\
{\it Fakult\"at f\"ur Physik, Universit\"at Bielefeld, D-33501 Bielefeld, Germany}}

\begin{abstract}
Shear viscosity of the gluon plasma in SU(3) YM theory is calculated nonperturbatively, within the stochastic
vacuum model. The result for the ratio of the shear viscosity to the entropy density, proportional to the squared chromo-magnetic gluon condensate and the 
fifth power of the correlation length of the chromo-magnetic vacuum, falls off with the increase of temperature. At temperatures larger than the deconfinement critical temperature by a factor of 2, this fall-off is determined 
by the sixth power of the temperature-dependent strong-coupling constant and yields an asymptotic 
approach to the conjectured lower bound of $1/(4\pi)$, achievable in ${\cal N}=4$ SYM theory. As a by-product of the calculation, we find a particular form of the two-point correlation function of gluonic field strengths, which is the only one consistent with the Lorentzian shape of the shear-viscosity spectral function. 
\end{abstract}

\maketitle

\section{Introduction and preliminary estimates}
The RHIC data on collective expansion dynamics of the quark-gluon plasma can be described by the relativistic hydrodynamics applied to a system with very large initial pressure gradients~\cite{ro}. According to these data, particles of different mass are emitted from the quark-gluon-plasma fireball with a common fluid velocity, that is a signature of a hydrodynamic expansion. Due to a large elliptic flow in noncentral collisions~\cite{odin}, 
an agreement between the experimental data~\cite{odin, dva} and the predictions of 
relativistic hydrodynamics can only be reached when the flow of the QGP-fluid is treated as almost non-viscous. 
This leads to
an indication that, in the vicinity of the deconfinement phase transition, 
the quark-gluon plasma produced in the RHIC experiments behaves more like an ideal quantum liquid rather than a weakly interacting gas. The mean free path $L_{\rm mfp}$ of a parton, which traverses such a liquid, is much smaller than the inter-particle distance, which is of the order of the inverse temperature $\beta=1/T$, i.e. 
$(L_{\rm mfp}^{\rm liq}/\beta)\ll 1$. 

One can consider for comparison a weakly interacting dilute-gas model of the 
quark-gluon plasma. There 
$L_{\rm mfp}^{\rm gas}\sim (\rho\sigma_t)^{-1}$ with $\rho$ and $\sigma_t$ standing for the particle-number 
density and the Coulomb transport cross section, respectively. Using the standard estimates $\rho\sim T^3$ and 
$\sigma_t\sim g^4\beta^2\ln g^{-1}$, where $g=g(T)$ is the perturbative finite-temperature QCD coupling, one obtains
$(L_{\rm mfp}^{\rm gas}/\beta)\sim 1/(g^4\ln g^{-1})\gg 1$, that strongly contradicts the above-mentioned
experimental results. One can check~\cite{mg} that these results could have only 
been reproduced by the dilute-gas model if
the perturbative transport cross section, $\sigma_t$, were larger by an order of magnitude.
This inconsistency of the weakly interacting quark-gluon plasma with the RHIC data initiated recent calculations
of kinetic coefficients in the {\it strongly} interacting relativistic plasmas. 

Among these coefficients, the one whose values define whether the plasma can be considered as weakly or strongly interacting is the shear viscosity $\eta$. It is related to the above $(L_{\rm mfp}/\beta)$-ratio via the
estimate $(\eta/s)\sim(L_{\rm mfp}/\beta)$, where $s$ is the entropy density.
According to this relation, the shear-viscosity to the entropy-density ratio, $\eta/s$,
becomes smaller when the plasma interacts stronger. For instance, for $T\sim 200{\,}{\rm MeV}$ and 
the estimated typical mean free path 
$L_{\rm mfp}\sim 0.1{\,}{\rm fm}$, one has $(\eta/s)\sim 0.1$. On the other hand, since the mean momentum change $\Delta p$
of a parton, which propagates through the plasma over the distance $L_{\rm mfp}$, is of the order of $T$, the Heisenberg uncertainty principle forbids the ratio $(L_{\rm mfp}/\beta)\sim L_{\rm mfp}\cdot\Delta p$ (and therefore also $\eta/s$) to be vanishingly small.
Up to now, the minimal value of $1/(4\pi)\simeq0.08$ for the shear-viscosity to the entropy-density ratio has been found in
${\cal N}=4$ SYM theory~\cite{pss}. It is thus challenging to find other QCD-motivated models where this ratio
would be that small. Recently, it has been demonstrated~\cite{fr} that such low values of the $(\eta/s)$-ratio can 
take place even in the perturbative YM plasma, due to the bremsstrahlung $gg\leftrightarrow ggg$ processes.
In Ref.~\cite{abm} it has been argued, though, that the perturbatively calculated collisional viscosity 
is anyhow larger (and therefore subdominant) compared to the so-called anomalous viscosity, which is generated by plasma instabilities. 

In this paper, we calculate the $(\eta/s)$-ratio in the gluon plasma of SU(3) YM theory nonperturbatively. 
We obtain the shear viscosity from the 
Kubo formula, which relates the corresponding spectral density $\rho(\omega)$ to the 
Euclidean correlation function of the $(1,2)$-component of the energy-momentum tensor $T_{12}({\bf x},x_4)$.
This method, proposed in Ref.~\cite{kw}, has been explored in
Refs.~\cite{ga, me} with the aim to simulate shear viscosity on the lattice.
Here we work in the continuum limit and parametrize the Euclidean correlation function of the energy-momentum tensors
by means of the stochastic vacuum model~\cite{ds}. This model generalizes QCD sum rules by assuming 
the existence of not only the gluon condensate $\left<g^2(F_{\mu\nu}^a)^2\right>$ but also of the finite 
vacuum correlation length $\mu^{-1}$. This assumption is justified by the lattice results on the exponential
fall-off at large distances of the two-point correlation function of gluonic field strengths~\cite{dmp, na},
$\left<F_{\mu\nu}^a(x)F_{\lambda\rho}^b(0)\right>\sim {\rm e}^{-\mu|x|}$. By virtue of this finding, the 
model manages to quantitatively describe confinement; for instance, the string tension reads 
$\sigma\propto \mu^{-2}\left<g^2(F_{\mu\nu}^a)^2\right>$. 

Below we will use a finite-temperature generalization of the 
stochastic vacuum model, accessible by implementing the Euclidean 
periodicity of the $x_4$-coordinate. In the deconfinement phase of interest, 
such a generalization yields for the spatial string tension $\sigma_s(T)$ a formula~\cite{aga} pretty similar to its above-quoted vacuum counterpart. This formula reads $\sigma_s(T)\propto \mu^{-2}(T)\left<g^2(F_{ij}^a)^2\right>_T$, where 
$\mu^{-1}(T)$ is the correlation length of the chromo-magnetic vacuum, and 
$\left<g^2(F_{ij}^a)^2\right>_T$ is the chromo-magnetic gluon condensate, which survives the deconfinement 
phase transition. The temperature dependence of the two main ingredients of the model, $\mu(T)$ and $\left<g^2(F_{ij}^a)^2\right>_T$, can be extracted from the results of the lattice simulations~\cite{f1, dmp}.

Since $T_{12}=g^2F_{1\mu}^a F_{2\mu}^a$, one {\it a~priori} expects from the Kubo formula, where the $\left<T_{12}(0)T_{12}(x)\right>$-correlator enters, 
that the shear viscosity
$\eta\propto \left<g^2(F_{ij}^a)^2\right>_T^2$. This is a general prediction of the stochastic vacuum model 
for all the 
kinetic coefficients, for example for the jet quenching parameter~\cite{ssp1}. In fact, according to 
the Kubo formula, all the kinetic coefficients are proportional to the total scattering
cross section of the propagating parton,  
which itself is proportional to 
$\left<g^2(F_{ij}^a)^2\right>_T^2$ in this model~\cite{cd, ssp}. Therefore, since 
the shear viscosity has the dimensionality of [mass]${}^3$, 
one can on entirely dimensional grounds expect for it the following result:
$$\eta\propto \mu^{-5}(T)\left<g^2(F_{ij}^a)^2\right>_T^2.$$
At temperatures larger than the temperature of dimensional reduction, $T>T_{*}$, 
$\mu(T)$ and $\left<g^2(F_{ij}^a)^2\right>_T^2$ 
are proportional to the corresponding power of the only 
dimensionful parameter present in the YM action at such temperatures, $g^2T$, i.e. 
$$\mu(T)\propto g^2T,~~~~ 
\left<g^2(F_{ij}^a)^2\right>_T\propto (g^2T)^4.$$
On the other hand, the entropy density $s(T)\propto T^3$, so that 
$$
\frac{\eta}{s}\propto g^6(T)~~~~ {\rm at}~~~~ T>T_{*}.$$
In this paper, we quantitatively answer the naturally arising  question of whether or not this function manages to get below the $(4\pi)^{-1}$-threshold at temperatures $T\lesssim 5T_c$, which are accessible experimentally and on the lattice.

The outline of the paper is as follows. In Section~II, by assuming an exponential fall-off for the two-point correlation 
function of the energy-momentum tensors $\left<T_{12}(0)T_{12}(x)\right>$, we obtain from 
the Kubo formula an integral equation for the 
spectral density $\rho(\omega)$ of $\eta$. In Section~III, by using for $\rho(\omega)$ a Lorentzian-type {\it ansatz}, 
with the same correlation length as that of $\left<T_{12}(0)T_{12}(x)\right>$, we explore this equation 
for the cases of large and small $|k|$'s, where $k$ is the number of a Matsubara mode.
The solution in the large-$|k|$ limit yields the range of variation of the numerical parameter $\alpha$, which enters the initial parametrization of $\left<T_{12}(0)T_{12}(x)\right>$. The solution in the small-$|k|$ limit can only coincide 
with the large-$|k|$ solution for a single value of $\alpha$ from this range. This fixes $\alpha$ completely and 
makes further calculations straightforward. In Section~IV, we analytically calculate the shear viscosity $\eta$. In Section~V, we use this result 
for $\eta$ to numerically find the ratio $\eta/s$. Also in Section~V we compare the calculated nonperturbative 
spectral density $\rho(\omega)$ with the perturbative one, which dominates at large $\omega$'s.
In Section~VI, we discuss the results of the paper, as well as possible further developments. 
In Appendix~A, we illustrate the separation of perturbative contributions to the Kubo formula from the nonperturbative ones.

\section{Kubo formula for the spectral density}

\noindent
Shear viscosity $\eta$ can be defined through the relation 
\begin{equation}
\label{eta} 
\left.\eta=\pi\frac{d \rho}{d \omega}\right|_{\omega=0},
\end{equation}
where the spectral density $\rho(\omega)$ is a solution to the following integral equation, called 
Kubo formula~\cite{kw, me}
\begin{equation}
\label{1}
\int_0^\infty d\omega\rho(\omega)\frac{\cosh\left[\omega\left(x_4-\frac{\beta}{2}\right)\right]}{\sinh(\omega\beta/2)}=
\int d^3x\sum\limits_{n=-\infty}^{+\infty}\left<T_{12}(0)T_{12}({\bf x},x_4-\beta n)\right>.
\end{equation}
Here the sum on the RHS runs over winding modes. We use the Fourier decomposition on the LHS of Eq.~(\ref{1}):
\begin{equation}
\label{2}
\frac{\cosh\left[\omega\left(x_4-\frac{\beta}{2}\right)\right]}{\sinh(\omega\beta/2)}=
2T\cdot \omega\sum\limits_{k=-\infty}^{+\infty}\frac{{\rm e}^{i\omega_kx_4}}{\omega^2+\omega_k^2},
\end{equation}
where $\omega_k=2\pi Tk$ is the $k$-th Matsubara frequency. The idea is to have a similar decomposition also on the 
RHS of Eq.~(\ref{1}). For the implementation of this idea, 
the following chain of equalities is important:
$$\sum\limits_{k=-\infty}^{+\infty}\frac{{\rm e}^{i\omega_kx_4}}{\left[1+(\omega_k/M)^2\right]^\alpha}=
\frac{1}{\Gamma(\alpha)}\sum\limits_{k=-\infty}^{+\infty}\int_0^\infty d\lambda \lambda^{\alpha-1}
{\rm e}^{-\lambda\left[1+(\omega_k/M)^2\right]+i\omega_kx_4}=$$
$$=\frac{M\beta}{2\sqrt{\pi}\Gamma(\alpha)}\int_0^\infty d\lambda \lambda^{\alpha-\frac32}{\rm e}^{-\lambda}
\sum\limits_{n=-\infty}^{+\infty}{\rm e}^{-\frac{M^2(x_4-\beta n)^2}{4\lambda}}=$$
$$=\frac{M^4\beta}{16\pi^2\Gamma(\alpha)}\int_0^\infty d\lambda \lambda^{\alpha-3}{\rm e}^{-\lambda}\int d^3x
\sum\limits_{n=-\infty}^{+\infty}{\rm e}^{-\frac{M^2\left[{\bf x}^2+(x_4-\beta n)^2\right]}{4\lambda}}=$$
$$=\frac{M^4\beta}{2^{\alpha+1}\pi^2\Gamma(\alpha)}\int d^3x
\sum\limits_{n=-\infty}^{+\infty}\frac{K_{2-\alpha}\left(M\sqrt{{\bf x}^2+(x_4-\beta n)^2}
\right)}{\left[M\sqrt{{\bf x}^2+(x_4-\beta n)^2}\right]^{2-\alpha}}.$$
Here $\alpha>0$ is some parameter, ``$\Gamma$'' and ``$K_{2-\alpha}$'' stand for the Gamma and the MacDonald functions, respectively. We assume that, at $T=0$,
\begin{equation}
\label{T12}
\left<T_{12}(0)T_{12}(x)\right>=N(\alpha)\left<G^2\right>^2\frac{K_{2-\alpha}(M|x|)}{(M|x|)^{2-\alpha}},
\end{equation}
where $\left<G^2\right>\equiv\left<g^2(F_{\mu\nu}^a)^2\right>$ and $N(\alpha)>0$ is a numerical coefficient, 
which will be determined. Then, in the deconfinement phase ($T>T_c$) of interest, the above chain of equalities yields
$${\rm RHS}~~ {\rm of}~~ {\rm Eq.}~~ (2)~~ = 2T\cdot \pi^2 2^\alpha\Gamma(\alpha)N(\alpha)
\left<G^2\right>_T^2 M^{2\alpha-4}
\sum\limits_{k=-\infty}^{+\infty}\frac{{\rm e}^{i\omega_kx_4}}{(\omega_k^2+M^2)^\alpha},$$
where $\left<G^2\right>_T\equiv\left<g^2(F_{ij}^a)^2\right>_T$. Note that,
for temperatures $T>T_{*}$, only the $(k=0)$-term in the sum should be considered,
since the dimensionally reduced theory is a theory of the zeroth Matsubara mode.

By using also Eq.~(\ref{2}), we can now rewrite 
Eq.~(\ref{1}) in terms of Fourier modes as
\begin{equation}
\label{om}
\int_0^\infty d\omega\rho(\omega)\frac{\omega}{\omega^2+\omega_k^2}=\pi^2 2^\alpha\Gamma(\alpha)N(\alpha)
\left<G^2\right>_T^2
\frac{M^{2\alpha-4}}{(\omega_k^2+M^2)^\alpha}.
\end{equation}
To solve this equation, we use the parametrization
$$\rho(\omega)=\omega f(\omega),$$ 
where $f(\omega)$ is some even function sufficient for the convergence of the 
integral at large $\omega$'s. 
Motivated by earlier works~\cite{kw, me, zeta}, we choose it in a Lorentzian-type form
\begin{equation}
\label{param}
f(\omega)=\frac{C}{(\omega^2+M^2)^{\alpha+\frac12}}.
\end{equation}
Here $C=C(T)>0$ is some function, which will be determined.
Apparently, the $\omega$-integration in Eq.~(\ref{om}) converges for any choice of 
$$\alpha>0.$$
Parametrization of the spectral density through Eq.~(\ref{param}) guarantees furthermore that both sides 
of Eq.~(\ref{om}) have the same leading large-$|k|$ behavior. It also implies~\cite{zeta} that 
$M=M(T)$ is the momentum scale below which perturbation theory breaks down. For this reason,
$M(T)$ should be of the order of the inverse 
correlation length of the chromo-magnetic vacuum, $\mu(T)$.
Shear viscosity can finally be obtained by means of Eq.~(\ref{eta}) as
\begin{equation}
\label{et1}
\eta=\frac{\pi C}{M^{2\alpha+1}}.
\end{equation}
We now solve Eq.~(\ref{om}) subsequently in the following cases:  $|k|\gg 1$ and $|k|\sim 1$.

\section{Contributions to $\eta$ from high and low Matsubara modes}

\subsection{$(|k|\gg 1)$-case}

\noindent
Plugging the Lorentzian-type {\it ansatz}~(\ref{param}) into Eq.~(\ref{om}), we obtain
$$
{\rm LHS}~ {\rm of}~ {\rm Eq.}~ (5)~ = C\int_0^\infty
d\omega\frac{\omega^2}{(\omega^2+\omega_k^2)(\omega^2+M^2)^{\alpha+\frac12}}=$$
$$=\frac{C}{\omega_k^{2\alpha}}
\int_0^\infty dx\frac{x^2}{(x^2+1)(x^2+\xi_k^2)^{\alpha+\frac12}},$$
where $x\equiv\omega/\omega_k$ and $\xi_k\equiv M/\omega_k$. At $|k|\gg 1$, one has 
\begin{equation}
\label{con1}
|\xi_k|\ll 1,
\end{equation}
and the latter integral yields
\begin{equation}
\label{LHS}
{\rm LHS}~ {\rm of}~ {\rm Eq.}~ (5)~
=\frac{C}{\omega_k^{2\alpha}}
\left\{\frac{1}{\xi_k^{2\alpha}}\left[\frac{\sqrt{\pi}
\Gamma(\alpha-1)}{4\Gamma(\alpha+\frac12)}\xi_k^2+{\cal O}(\xi_k^3)\right]
+\frac{\pi}{2\sin(\pi\alpha)}+{\cal O}(\xi_k^2)\right\}.
\end{equation}
We see that, if $\alpha<1$, then the 
leading term of the expansion is $\propto\frac{\pi}{2\sin(\pi\alpha)}$, i.e. it is $k$-independent, whereas otherwise 
the leading term of the expansion becomes $k$-dependent. For this reason, we restrict ourselves to 
$$\alpha<1$$ 
only. Furthermore, since $\alpha\xi_k^2\ll 1$ as well, 
one can expand in powers of $\xi_k$ also the RHS of Eq.~(\ref{om}) to obtain
\begin{equation}
\label{RHS}
{\rm RHS}~ {\rm of}~ {\rm Eq.}~ (5)~~ = \pi^2 2^\alpha\Gamma(\alpha)N(\alpha)\frac{\left<G^2\right>_T^2}{\omega_k^{2\alpha}}
M^{2\alpha-4}[1+{\cal O}(\xi_k^2)].
\end{equation} 
Equations~(\ref{LHS}) and (\ref{RHS}) yield the function $C$:
$$C=\pi 2^{\alpha+1}\Gamma(\alpha)N(\alpha)\sin(\pi\alpha)\left<G^2\right>_T^2 M^{2\alpha-4}.$$
Accordingly, Eq.~(\ref{et1}) yields for the shear viscosity 
\begin{equation}
\label{gg1}
\eta\Bigr|_{|k|\gg 1}=
\pi^2 2^{\alpha+1}\Gamma(\alpha)N(\alpha)\sin(\pi\alpha)\frac{\left<G^2\right>_T^2}{M^5}.
\end{equation}
The parametric dependence of this expression on $\left<G^2\right>_T$ and $M$ is indeed the one following from the 
elementary dimensional analysis made in Introduction.

\subsection{$(|k|\sim 1)$-case}

\noindent
Consider $|k|$'s sufficiently small for the inequality 
\begin{equation}
\label{sma}
|\omega_k|\ll M,~~~ {\rm i.e.}~~~ |\xi_k|\gg 1,
\end{equation} 
to hold. Disregarding terms ${\cal O}(\omega_k^2/M^2)$ and higher, one has
$${\rm LHS}~ {\rm of}~ {\rm Eq.}~ (5)~~ \simeq 
\frac{C}{M^{2\alpha}}\int_0^\infty\frac{dz}{(z^2+1)^{\alpha+\frac12}}=\frac{\sqrt{\pi}}{2}
\frac{\Gamma(\alpha)}{\Gamma\bigl(\alpha+\frac12\bigr)}\frac{C}{M^{2\alpha}},$$
where $z\equiv\omega/M$, while
$${\rm RHS}~ {\rm of}~ {\rm Eq.}~ (5)~~ \simeq \pi^2 2^\alpha\Gamma(\alpha)N(\alpha)
\frac{\left<G^2\right>_T^2}{M^4}.$$
We obtain from these two equations
$$C=\pi^{3/2}2^{\alpha+1}\Gamma\Bigl(\alpha+\frac12\Bigr)N(\alpha)\left<G^2\right>_T^2 M^{2\alpha-4}$$
and, according to Eq.~(\ref{et1}), 
\begin{equation}
\label{eq0}
\eta\Bigr|_{|k|\sim 1}=\pi^{5/2}2^{\alpha+1}\Gamma\Bigl(\alpha+\frac12\Bigr)N(\alpha)
\frac{\left<G^2\right>_T^2}{M^5}.
\end{equation}
In particular, at $T>T_{*}$, where only the $(k=0)$-mode should be considered, this result becomes exact.

\begin{figure}
\psfrag{P}{$\eta\bigr|_{|k|\gg 1}{\,}/{\,}\eta\bigr|_{|k|\sim 1}$}
\psfrag{A}{\Large{$\alpha$}}
\epsfig{file=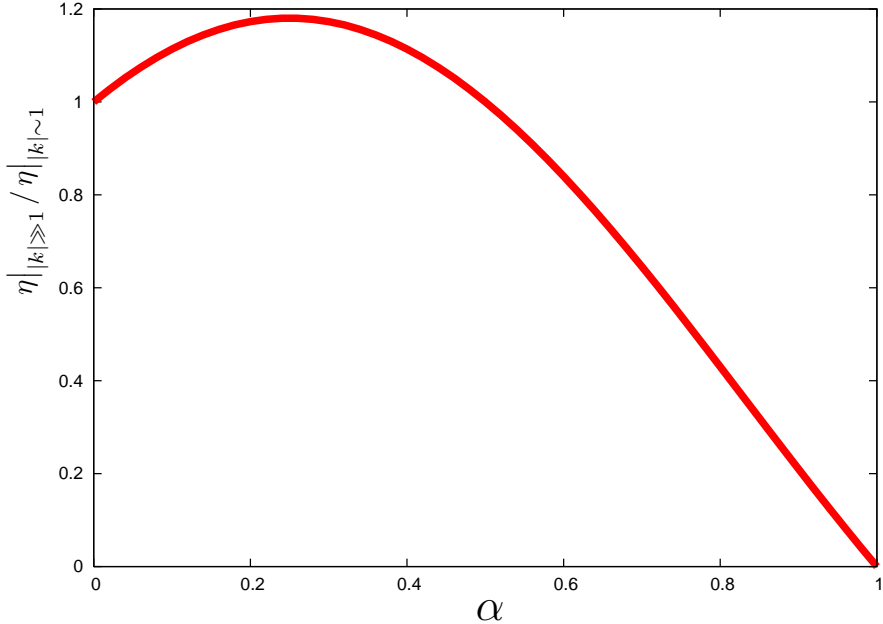, width=120mm}
\caption{\label{111}The ratio $\frac{\eta\bigr|_{|k|\gg 1}}{\eta\bigr|_{|k|\sim 1}}$ given by Eq.~(\ref{ratio}).}
\end{figure}

\section{$\eta$ from the correlation function $\left<T_{12}(0)T_{12}(x)\right>$}

\noindent
We determine now the parameters $N(\alpha)$ and $M$, which enter the correlation function~(\ref{T12}). 
This correlation function reads
$$\left<T_{12}(0)T_{12}(x)\right>=\left<g^4F_{1\mu}^a(0)F_{2\mu}^a(0)F_{1\nu}^b(x)F_{2\nu}^b(x)\right>=$$
$$=\left<\left<g^4F_{1\mu}^a(0)F_{2\mu}^a(0)F_{1\nu}^b(x)F_{2\nu}^b(x)\right>\right>+
\left<g^2F_{1\mu}^a(0)F_{2\mu}^a(0)\right>\left<g^2F_{1\nu}^b(x)F_{2\nu}^b(x)\right>+$$
\begin{equation}
\label{TT}
+\left<g^2F_{1\mu}^a(0)F_{1\nu}^b(x)\right>\left<g^2F_{2\mu}^a(0)F_{2\nu}^b(x)\right>+
\left<g^2F_{1\mu}^a(0)F_{2\nu}^b(x)\right>\left<g^2F_{2\mu}^a(0)F_{1\nu}^b(x)\right>,
\end{equation}
where double angular brackets denote a connected (or irreducible) average. We use the Gaussian-dominance 
hypothesis~\cite{ds}, which allows one to disregard this connected average.
For the two-point correlation function of gluonic field strengths
we use the standard parametrization~\cite{ds, ssp}
\begin{equation}
\label{sta}
\left<g^2F_{\mu\nu}^a(x)F_{\lambda\rho}^b(x')\right>=\left(\delta_{\mu\lambda}\delta_{\nu\rho}-
\delta_{\mu\rho}\delta_{\nu\lambda}\right)
\frac{\left<G^2\right>}{12(N_c^2-1)}\delta^{ab}D(x-x'),
\end{equation}
where $D(x)$ is a dimensionless function mediating the confining interaction.
In this parametrization, we have disregarded 
a small contribution of non-confining non-perturbative interactions~\cite{ssp1},~\footnote{High-energy scattering data in the vacuum suggest that the relative weight of these interactions constitutes of at most 26\%~\cite{ssp}.}.
By using Eq.~(\ref{sta}), we obtain for the correlation function~(\ref{TT}):
$$\left<T_{12}(0)T_{12}(x)\right>\simeq
\left<g^2F_{1\mu}^a(0)F_{2\mu}^a(0)\right>\left<g^2F_{1\nu}^b(x)F_{2\nu}^b(x)\right>+
\left<g^2F_{1\mu}^a(0)F_{1\nu}^b(x)\right>\left<g^2F_{2\mu}^a(0)F_{2\nu}^b(x)\right>+$$
\begin{equation}
\label{disc} 
+\left<g^2F_{1\mu}^a(0)F_{2\nu}^b(x)\right>\left<g^2F_{2\mu}^a(0)F_{1\nu}^b(x)\right>
=\frac{\left<G^2\right>^2}{72(N_c^2-1)}D^2(x).
\end{equation}
The dimensionless function $D(x)$ is usually chosen in the form
\begin{equation}
\label{d}
D(x)={\rm e}^{-\mu|x|}.
\end{equation}
Inserting this expression into the formula for the string tension 
in the fundamental representation,
\begin{equation}
\label{sig}
\sigma_{\rm f}=\frac{\left<G^2\right>}{144}\int d^2x D(x),
\end{equation}
one can define the gluon condensate in terms of $\sigma_{\rm f}$ and the vacuum correlation length 
$\mu$ as follows~\cite{ssp,ssp1}: 
\begin{equation}
\label{glc}
\left<G^2\right>=\frac{72}{\pi}\sigma_{\rm f}\mu^2.
\end{equation}

To obtain for the correlator $\left<T_{12}(0)T_{12}(x)\right>$ the functional form given by the 
RHS of Eq.~(\ref{T12}), we modify parametrization~(\ref{d}) to
\begin{equation}
\label{modD}
D(x)={\cal A}(\alpha)\sqrt{\frac{K_{2-\alpha}(2\mu|x|)}{(2\mu|x|)^{2-\alpha}}},
\end{equation}
where ${\cal A}(\alpha)$ is a numerical normalization factor. At $|x|\gtrsim\mu^{-1}$, the new function~(\ref{modD}) 
falls off with the same exponent as Eq.~(\ref{d}).
To obtain the normalization factor ${\cal A}(\alpha)$, we substitute Eq.~(\ref{modD}) into relation~(\ref{sig}), 
which holds for any function $D(x)$.
Using further expression~(\ref{glc}), we obtain
\begin{equation}
\label{CalA}
{\cal A}(\alpha)=\frac{4}{\int_0^\infty dz \sqrt{z^\alpha K_{2-\alpha}(z)}}.
\end{equation}
The correlator~(\ref{disc}) now reads
\begin{equation}
\label{t123}
\left<T_{12}(0)T_{12}(x)\right>=\frac{{\cal A}^2(\alpha)}{576}
\left<G^2\right>^2
\frac{K_{2-\alpha}(2\mu|x|)}{(2\mu|x|)^{2-\alpha}},
\end{equation}
where the function ${\cal A}(\alpha)$ is given by Eq.~(\ref{CalA}), and we have fixed $N_c=3$.
Comparing Eq.~(\ref{t123}) with the original definition~(\ref{T12}), we conclude that
$$N(\alpha)=\frac{{\cal A}^2(\alpha)}{576}~~~ {\rm and}~~~ M=2\mu.$$
Equations~(\ref{gg1}) and (\ref{eq0}) yield now contributions to $\eta$ from high and low Matsubara modes:
\begin{equation}
\label{eta1}
\eta\Bigr|_{|k|\gg 1}=\frac{\pi^2}{9216}2^\alpha\Gamma(\alpha)\sin(\pi\alpha)
\frac{[{\cal A}(\alpha)\left<G^2\right>_T]^2}{\mu^5(T)}
\end{equation}
and
\begin{equation}
\label{eta2}
\eta\Bigr|_{|k|\sim 1}=\frac{\pi^{5/2}}{9216}2^\alpha\Gamma\Bigl(\alpha+\frac12\Bigr)
\frac{[{\cal A}(\alpha)\left<G^2\right>_T]^2}{\mu^5(T)}.
\end{equation}
The ratio of these results,
\begin{equation}
\label{ratio}
\frac{\eta\Bigr|_{|k|\gg 1}}{\eta\Bigr|_{|k|\sim 1}}=
\frac{\Gamma(\alpha)\sin(\pi\alpha)}{\sqrt{\pi}\Gamma\Bigl(\alpha+\frac12\Bigr)},
\end{equation}
in the interval $0<\alpha<1$ of interest is plotted in Fig.~\ref{111}. It equals to unity at $\alpha=1/2$, i.e., at this  value of $\alpha$, our results for shear viscosity become 
$k$-independent, as they should be.
This yields the principal analytic result of the present paper:
\begin{equation}
\label{etA}
\eta(T)=\frac{\pi^{5/2}}{4608\sqrt{2}}\frac{[{\cal A}(1/2)\left<G^2\right>_T]^2}{\mu^5(T)},
\end{equation}
where ${\cal A}(1/2)\simeq1.05$. Remarkably, {\it ansatz}~(\ref{param}) at $\alpha=1/2$ takes the conventional 
Lorentzian form. 
In the next Section, we will evaluate the ratio $\eta/s$ numerically.

\begin{figure}
\psfrag{P}{$s(T)/T^3$}
\epsfig{file=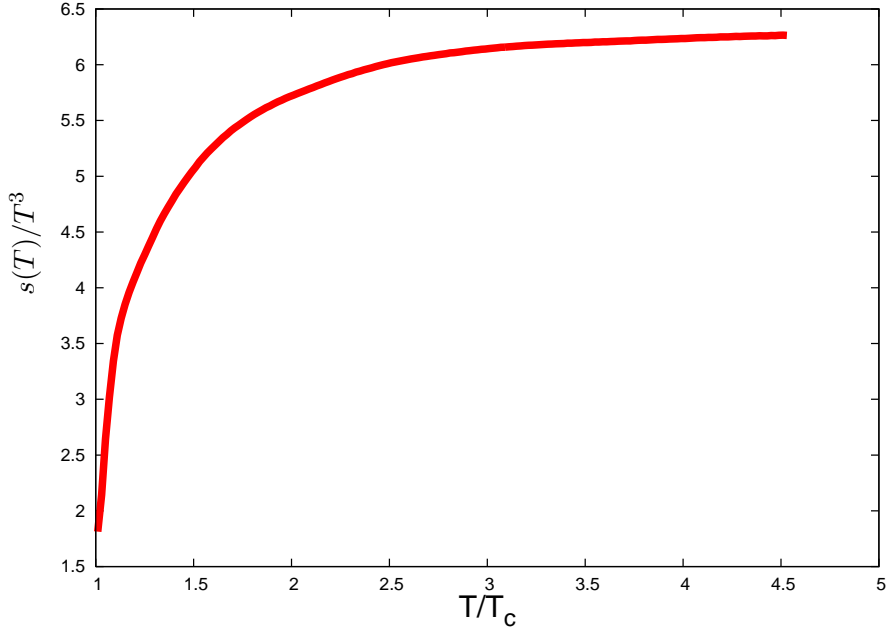, width=120mm}
\caption{\label{222}Entropy density $s(T)$ in the units of $T^3$ obtained from the lattice values 
for the pressure $p_{\rm lat}$~\cite{f1} (courtesy of F.~Karsch).}
\end{figure}

\section{Numerical evaluation}

\noindent
Following Ref.~\cite{f1}, we assume the value $T_c=270{\,}{\rm MeV}$ in SU(3) YM theory. We use
the two-loop running coupling~\cite{f1} 
$$g^{-2}(T)=2b_0\ln\frac{T}{\Lambda}+\frac{b_1}{b_0}\ln\left(2\ln\frac{T}{\Lambda}\right),~ {\rm where}~
b_0=\frac{11N_c}{48\pi^2},~ b_1=\frac{34}{3}\left(\frac{N_c}{16\pi^2}\right)^2,~ \Lambda=0.104T_c,
$$
and $N_c=3$ for the case under study.
We also assume for $\mu(T)$ and for the spatial string tension in the fundamental representation, $\sigma_{\rm f}(T)$,
the following parametrizations~\cite{aga, ssp1}:
\begin{equation}
\label{MU}
\mu(T)=\mu\cdot\left\{\begin{array}{rcl}1~~ {\rm at}~~ T_c<T<T_{*},\\
\frac{g^2(T)T}{g^2(T_{\rm d.r.})T_{\rm d.r.}}~~ {\rm at}~~ T>T_{*},\end{array}\right.
\end{equation}
\begin{equation}
\label{SI}
\sigma_{\rm f}(T)=\sigma_{\rm f}\cdot\left\{\begin{array}{rcl}1~~ {\rm at}~~ T_c<T<T_{*},\\
\left[\frac{g^2(T)T}{g^2(T_{\rm d.r.})T_{\rm d.r.}}\right]^2~~ {\rm at}~~ T>T_{*},\end{array}\right.
\end{equation}
where $\mu=894{\,}{\rm MeV}$~\cite{dmp} and $\sigma_{\rm f}=(0.44{\,}{\rm GeV})^2$. Equation~(\ref{glc}), extrapolated to
finite temperatures, yields for the chromo-magnetic gluon condensate $\left<G^2\right>_T$~\cite{aga, ssp1}:
$$\left<G^2\right>_T=\frac{72}{\pi}\sigma_{\rm f}(T)\mu^2(T)=\left<G^2\right>
\cdot\left\{\begin{array}{rcl}1~~ {\rm at}~~ T_c<T<T_{*},\\
\left[\frac{g^2(T)T}{g^2(T_{\rm d.r.})T_{\rm d.r.}}\right]^4~~ {\rm at}~~ T>T_{*}.\end{array}\right.$$
The value of $T_{*}$ can be estimated from the equation
$$\sigma_{\rm f}(T_{*})=\sigma_{\rm f},$$ 
where $\sigma_{\rm f}(T)=[0.566g^2(T)T]^2$ is 
the high-temperature parametrization of the fundamental spatial string tension~\cite{f1}.  
Solving this equation numerically, one obtains~\cite{aps}
$$T_{*}=1.28T_c.$$
The entropy density $s=s(T)$ can be obtained by the formula
$s=\partial p_{\rm lat}/\partial T$, where we use for the pressure $p_{\rm lat}$ 
the corresponding lattice values from Ref.~\cite{f1}. In Fig.~\ref{222}, we plot $s(T)$ in the units of $T^3$.
The temperature dependence of the ratio $\eta/s$ is determined by the function $\left<G^2\right>_T^2/[\mu^5(T)s(T)]$. One can check numerically that, at $T\gtrsim 2T_c$ where $s/T^3$ is nearly constant, 
$\left<G^2\right>_T^2/[\mu^5(T)s(T)]={\cal O}\bigl(g^6(T)\bigr)$, as was mentioned in Introduction.

\begin{figure}
\psfrag{P}{\Large{$\eta/s$}}
\epsfig{file=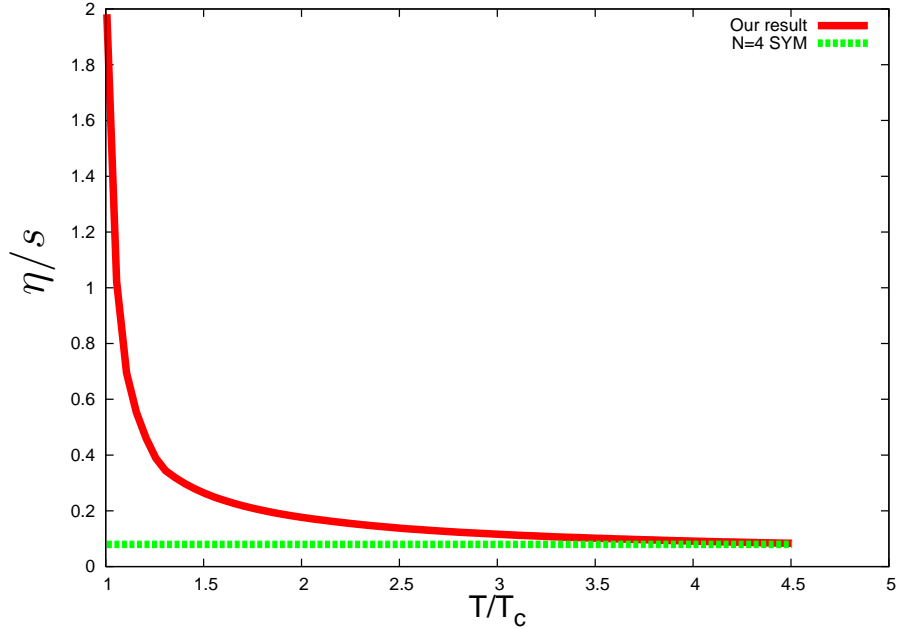, width=120mm}
\caption{\label{4}Calculated values of the ratio $\eta/s$ as a function of temperature. Also shown is the 
conjectured lower bound of $(4\pi)^{-1}$ for this quantity, realized in ${\cal N}=4$ SYM.}
\end{figure}

In Fig.~\ref{4}, we plot the ratio $\eta/s$, with $\eta$ given by Eq.~(\ref{etA}), as a function of temperature.
Also in Fig.~\ref{4}, we plot the conjectured 
lower bound for this ratio, equal to $(4\pi)^{-1}\simeq0.08$, which is realized in ${\cal N}=4$ SYM~\cite{pss}.
This bound is indeed not reached by our values, although they get very close to it at the highest temperature $T=4.54T_c$ where the lattice data for the pressure (and therefore also for $s$) are available.

\begin{figure}
\psfrag{y}{$\omega/T$}
\psfrag{z}{\Large{$\rho^{\rm full}(\omega, T)$}}
\epsfig{file=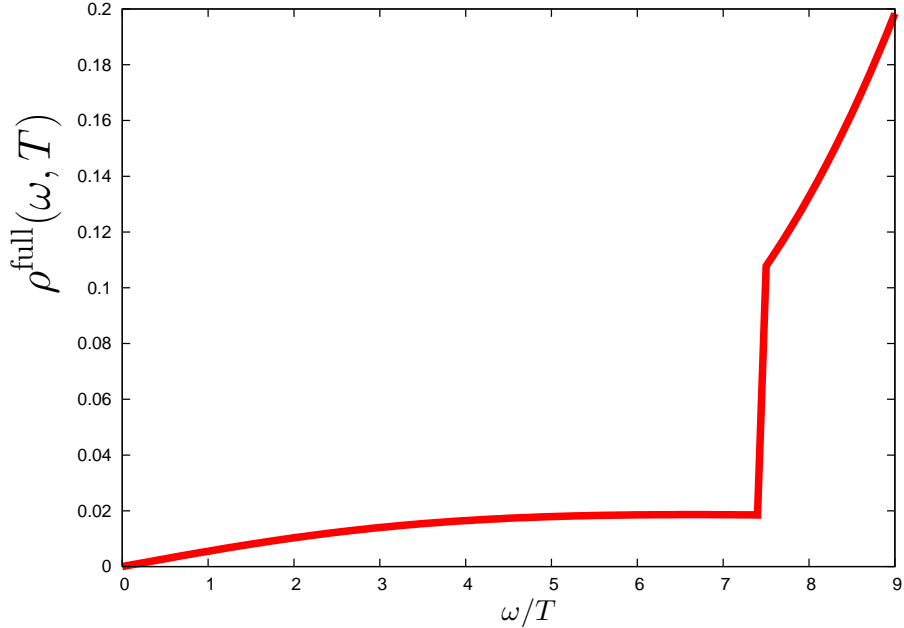, width=120mm}
\caption{\label{46}The full spectral density $\rho^{\rm full}(\omega, T)$ as a function of $\omega/T$ at $T=T_c$. 
At $\frac{\omega}{T}<7.5$, $\rho^{\rm full}$ is entirely nonperturbative and given by Eq.~(\ref{calc}), while 
at $\frac{\omega}{T}>7.5$ the perturbative part $\rho^{\rm pert}$ is dominating.}
\end{figure}

Furthermore, we compare numerically the obtained nonperturbative spectral density,
\begin{equation}
\label{calc}
\rho(\omega, T)=C(T){\,}\frac{\omega}{\omega^2+4\mu^2(T)},~~ {\rm where}~~ C(T)=\left(\frac{\pi}{2}\right)^{3/2}
\frac{{\cal A}^2(1/2)}{576}{\,}\frac{\left<G^2\right>_T^2}{\mu^3(T)},
\end{equation}
with the perturbative one, which at the tree level reads~\cite{ga, me}
$$\rho^{\rm pert}(\omega, T)=\frac{1}{20\pi^2}{\,}\theta(\omega-\tilde\omega(T)){\,}\frac{\omega^4}{\tanh\frac{\omega}{4T}}.$$
Note that $\rho^{\rm pert}(\omega, T)\ne 0$ only at $\omega>\tilde\omega(T)$, where $\tilde\omega(T)\simeq7.5T$~\cite{me}. 
For this reason, $\rho^{\rm pert}(\omega, T)$ in any case does not affect the calculated $\eta$, which is defined 
according to Eq.~(\ref{eta}) by the values of the spectral density at $\omega\to 0$. 
Figure~\ref{46} illustrates the full spectral density
$\rho^{\rm full}=\rho+\rho^{\rm pert}$ as a function of $\omega/T$ at $T=T_c$. At $\omega<\tilde\omega(T)$, 
$\rho^{\rm full}$
is given by the obtained result~(\ref{calc}), while at $\omega>\tilde\omega(T)$ 
the perturbative part $\rho^{\rm pert}$ takes it over.
Were $\rho^{\rm pert}$ nonvanishing down to $\omega=0$, it would dominate over $\rho$ already at 
$\omega>4.85T$. That is the reason why, by $\omega=\tilde\omega(T)$, $\rho^{\rm pert}$ 
significantly exceeds $\rho$, as one can see from the gap in the values of $\rho^{\rm full}$ at this value of $\omega$. Qualitatively, the same $\rho^{\rm full}$ and the relation between $\rho$ and $\rho^{\rm pert}$ persist with the increase of temperature. 

Finally, in Appendix~A, we illustrate the correspondence between the splitting of $\rho^{\rm full}=\rho+\rho^{\rm pert}$ and the splitting of $\left<T_{12}(0)T_{12}(x)\right>_{\rm full}=\left<T_{12}(0)T_{12}(x)\right>+ \left<T_{12}(0)T_{12}(x)\right>_{\rm pert}$.
This correspondence allows one to isolate the contribution, which $\rho^{\rm pert}$ brings about to the 
$\omega$-integral in the full Kubo formula.

\section{Discussion and outlook}

\noindent
In this paper, we have applied Kubo formula to a nonperturbative calculation of the shear viscosity $\eta$ 
in SU(3) YM theory. With the use of the stochastic vacuum model, 
the $\left<T_{12}(0)T_{12}(x)\right>$-correlator entering Kubo formula has been 
expressed in terms of the temperature-dependent chromo-magnetic gluon condensate $\left<g^2(F_{ij}^a)^2\right>_T$ 
and the correlation length of the chromo-magnetic vacuum $\mu^{-1}(T)$. As was expected (cf.~Introduction), 
$\eta$ turns out to be $\propto\mu^{-5}(T)\left<g^2(F_{ij}^a)^2\right>_T^2$, where the numerical factor 
is given by Eq.~(\ref{etA}). At temperatures $T\gtrsim 2T_c$, the calculated ratio $\eta/s$ falls off as 
${\cal O}\bigl(g^6(T)\bigr)$, as it should do in the dimensionally-reduced theory. 
Numerically, up to the 
temperature $T=4.54T_c$, where the lattice data on bulk thermodynamic quantities are still available, 
the obtained values of the ratio $\eta/s$ stay above the conjectured lower bound of $(4\pi)^{-1}$, which is reached  
in ${\cal N}=4$ SYM. 

Formally, our result~(\ref{etA}) persists even at higher temperatures, being extrapolated to which it
yields for the $(\eta/s)$-ratio values smaller than $(4\pi)^{-1}$. 
One should, however, realize that the monotonic fall-off of $\eta/s$ with temperature, stemming from the relation
$\eta\propto\mu^{-5}(T)\left<g^2(F_{ij}^a)^2\right>_T^2$, is predefined by our calculational method, which combines 
Kubo formula with the stochastic vacuum model. In fact,
all the kinetic coefficients derivable in this way should be $\propto\left<g^2(F_{ij}^a)^2\right>_T^2$ 
(cf.~Introduction). In particular, this is the case for the bulk viscosity $\zeta$~\cite{zeta}, which can be 
obtained from the correlation function
$$\left<g^4F_{\mu\nu}^{a{\,}2}(0)F_{\lambda\rho}^{b{\,}2}(x)\right>\simeq
\left<G^2\right>^2\left[1+\frac{1}{3(N_c^2-1)}D^2(x)\right].$$
(Here ``$\simeq$'' stands for ``Gaussian approximation''.)
On general grounds~\cite{kap}, one indeed expects a monotonic fall-off of the $(\zeta/s)$-ratio with temperature, as was 
confirmed by explicit calculations~\cite{zeta}. However, on the same general grounds~\cite{kap}, 
for the $(\eta/s)$-ratio in question one expects the existence of a minimum at temperatures close to 
$T_c$~\footnote{Such a minimum appears, for example, upon the multiplication of the perturbative result for the 
$(\eta/s)$-ratio~\cite{amy} by the squared fundamental Polyakov loop~\cite{PH}.} and a subsequent {\it increase} 
with the further increase of temperature. Indeed, at least at the temperature as high as $47.4T_c$, 
$g(T)$ becomes smaller than unity, and the weakly-interacting dilute-gas model
of the gluon plasma gradually sets in. As mentioned in 
Introduction, in the dilute-gas model~\cite{amy} $(\eta/s)\sim 1/(g^4\ln g^{-1})$, i.e. this ratio increases with temperature.  
The stochastic vacuum model, on the other hand, being applicable at strong coupling,
correctly yields the expected fall-off of the $(\eta/s)$-ratio at temperatures $\lesssim 2T_c$, 
but cannot reproduce its increase at much higher temperatures.

We would also like to emphasize an interesting fact, which has been realized by the end of the calculation.
We have started with the general $\alpha$-dependent Lorentzian-type {\it ansatz}~(\ref{param}) for the spectral 
density $\rho(\omega)$. By using it in the Kubo formula, we have come to the conclusion that only for the single
value, $\alpha=1/2$, this {\it ansatz} provides the Matsubara-mode independence of $\rho(\omega)$. For this value 
of $\alpha$, Eq.~(\ref{param}) takes the conventional Lorentzian form. In this way, also 
the function $D(x)$ from Eq.~(\ref{sta}) is defined unambiguously as 
$$D(x)={\cal A}(1/2){\,}\sqrt{\frac{K_{3/2}(2\mu|x|)}{(2\mu|x|)^{3/2}}},$$
where the value of ${\cal A}(1/2)$ can be found after Eq.~(\ref{etA}).

Note finally that we have used in our calculation the Gaussian-dominance hypothesis~\cite{ds}, which allows one 
to disregard in Eq.~(\ref{TT}) the connected four-point correlation function of gluonic field strengths 
compared to the pairwise products of the two-point correlation functions. The same approximation was
used in Ref.~\cite{mk} for the calculation of topological susceptibility via the four-point correlation function.
This approximation can be relaxed in the following way. Consider a parametrization for the 
nonperturbative part of the connected four-point correlation function suggested in Ref.~\cite{kd}:
$$\left<\left<g^4F_{\mu_1\nu_1}^{a_1}(x_1)F_{\mu_2\nu_2}^{a_2}(x_2)F_{\mu_3\nu_3}^{a_3}(x_3)F_{\mu_4\nu_4}^{a_4}(x_4)
\right>\right>=$$
$$=\left<G^2\right>^2\Bigl\{
f^{a_1a_2b}f^{a_3a_4b}\left(\varepsilon_{\mu_1\nu_1\mu_3\nu_3}\varepsilon_{\mu_2\nu_2\mu_4\nu_4}-
\varepsilon_{\mu_1\nu_1\mu_4\nu_4}\varepsilon_{\mu_2\nu_2\mu_3\nu_3}\right)+$$
$$+f^{a_3a_1b}f^{a_2a_4b}\left(\varepsilon_{\mu_1\nu_1\mu_4\nu_4}\varepsilon_{\mu_2\nu_2\mu_3\nu_3}-
\varepsilon_{\mu_1\nu_1\mu_2\nu_2}\varepsilon_{\mu_3\nu_3\mu_4\nu_4}\right)+$$
$$+f^{a_2a_3b}f^{a_1a_4b}\left(\varepsilon_{\mu_1\nu_1\mu_2\nu_2}\varepsilon_{\mu_3\nu_3\mu_4\nu_4}-
\varepsilon_{\mu_1\nu_1\mu_3\nu_3}\varepsilon_{\mu_2\nu_2\mu_4\nu_4}\right)-$$
$$-\frac18\Bigl[\delta^{a_1a_2}\delta^{a_3a_4}\left(\delta_{\mu_1\mu_2}\delta_{\nu_1\nu_2}-
\delta_{\mu_1\nu_2}\delta_{\mu_2\nu_1}\right)\left(\delta_{\mu_3\mu_4}\delta_{\nu_3\nu_4}-
\delta_{\mu_3\nu_4}\delta_{\mu_4\nu_3}\right)+$$
$$+\delta^{a_1a_3}\delta^{a_2a_4}\left(\delta_{\mu_1\mu_3}\delta_{\nu_1\nu_3}-
\delta_{\mu_1\nu_3}\delta_{\mu_3\nu_1}\right)\left(\delta_{\mu_2\mu_4}\delta_{\nu_2\nu_4}-
\delta_{\mu_2\nu_4}\delta_{\mu_4\nu_2}\right)+$$
\begin{equation}
\label{to}
+\delta^{a_1a_4}\delta^{a_2a_3}\left(\delta_{\mu_1\mu_4}\delta_{\nu_1\nu_4}-
\delta_{\mu_1\nu_4}\delta_{\mu_4\nu_1}\right)\left(\delta_{\mu_2\mu_3}\delta_{\nu_2\nu_3}-
\delta_{\mu_2\nu_3}\delta_{\mu_3\nu_2}\right)\Bigr]\Bigr\}\tilde D(z_1,\ldots, z_6),
\end{equation}
where $z_1=x_1-x_2$, $z_2=x_1-x_3$,..., $z_6=x_3-x_4$ are relative coordinates. For the connected 
correlation function entering Eq.~(\ref{TT}) this parametrization yields
$$
\left<\left<g^4F_{1\mu}^a(0)F_{2\mu}^a(0)F_{1\nu}^b(x)F_{2\nu}^b(x)\right>\right>=
2(N_c^2-1)\left(N_c-\frac18\right)\left<G^2\right>^2\tilde D(0,x,x,x,x,0).$$
Similarly to Eq.~(\ref{modD}), for the function $\tilde D$ too
one can have a parametrization compatible with Eq.~(\ref{T12}):
$$\tilde D(0,x,x,x,x,0)={\cal B}(\alpha)\frac{K_{2-\alpha}(4\mu|x|)}{(4\mu|x|)^{2-\alpha}}.$$
The normalization factor ${\cal B}(\alpha)$ should now be determined simultaneously with the 
normalization factor ${\cal A}(\alpha)$ from a system of equations for two observables, which both depend 
on the functions $D$ and $\tilde D$. Natural observables of this kind are the string tension and the 
topological susceptibility. The contribution of the function $\tilde D$ to the string tension has already been 
evaluated in Ref.~\cite{kd}. Further analysis of the outlined extension of the Gaussian approximation is, however, not the 
purpose of the present paper.

\acknowledgements

\noindent
I am grateful to Frithjof Karsch, Hans-J\"urgen Pirner and Arif Shoshi for helpful discussions, and to Yoshimasa Hidaka 
and Olaf Kaczmarek for the useful correspondence and comments. I also thank 
Frithjof Karsch for providing the details of the lattice data from Ref.~\cite{f1}. 
This work has been supported by the German Research Foundation (DFG), contract Sh~92/2-1.

\section*{Appendix A. Matching perturbative contributions in the Kubo formula.}

\noindent
For this Appendix, we promote Eq.~(\ref{1}) to the full Kubo formula, i.e. replace $\rho$ by 
$\rho^{\rm full}=\rho+\rho^{\rm pert}$
and $\left<T_{12}(0)T_{12}(x)\right>$ by $\left<T_{12}(0)T_{12}(x)\right>_{\rm full}=\left<T_{12}(0)T_{12}(x)\right>+
\left<T_{12}(0)T_{12}(x)\right>_{\rm pert}$. On the LHS of such a full Kubo formula, consider the integral containing
$\rho_{\rm pert}$.
To facilitate the $\omega$-integration, we approximate the perturbative part of the spectral density by the function 
$\rho^{\rm pert}=N\omega^4$ down to $\omega=0$ and determine the coefficient $N$. The integral emerging on the 
LHS of the Kubo formula then reads 
$$\int_0^\infty d\omega{\,}\omega^4{\,}\left[\cosh(\omega x_4)\coth(\omega\beta/2)-\sinh(\omega x_4)\right]=
\sum\limits_{n=0}^{\infty}\left[\frac{1}{(\beta n+x_4)^5}+\frac{1}{[\beta n+(\beta-x_4)]^5}\right]=$$
$$=24\left\{\frac{1}{x_4^5}+\sum\limits_{n=1}^{\infty}\left[\frac{1}{(\beta n+x_4)^5}+\frac{1}{(\beta n-x_4)^5}
\right]\right\}.\eqno(A.1)$$
To obtain the last expression, we have extracted the $(n=0)$-term from the sum
$\sum\limits_{n=0}^{\infty}\frac{1}{(\beta n+x_4)^5}$ and 
shifted $n$ by 1 in the sum $\sum\limits_{n=0}^{\infty}\frac{1}{[\beta n+(\beta-x_4)]^5}$.
Note that $x_4\in[0,\beta]$, and the obtained expression has singularities at $x_4=0$ and 
$x_4=\beta$ [in the $(n=1)$-term]. These singularities are identical, since 
$\cosh\left[\omega\left(x_4-\frac{\beta}{2}\right)\right]=\cosh(\omega\beta/2)$ both at $x_4=0$ and $x_4=\beta$.

We will demonstrate now that this expression corresponds to the contribution, which 
$\left<T_{12}(0)T_{12}(x)\right>_{\rm pert}$ brings about to the RHS of the full Kubo formula. The UV-finite part 
of this perturbative correlation function can be written as 
$$\left<T_{12}(0)T_{12}(x)\right>_{\rm pert}=\frac{A(N_c^2-1)g^4}{x^8},$$
where the value of the numerical constant $A$ depends on the regularization scheme applied.
Thus, the sum emerging on the RHS of the Kubo formula reads
$$\sum\limits_{n=-\infty}^{+\infty}\frac{1}{[{\bf x}^2+(\beta n+x_4)^2]^4}.$$
Doing the integration over $d^3x$ first, we have
$$\int_0^\infty dx\frac{x^2}{[x^2+(\beta n+x_4)^2]^4}=\frac{\pi}{32|\beta n+x_4|^5}.$$
The part of the sum with negative winding modes reads
$$\sum\limits_{n=-\infty}^{-1}\frac{1}{|\beta n+x_4|^5}=\sum\limits_{n=-\infty}^{-1}\frac{1}{(-\beta n-x_4)^5}=
\sum\limits_{n=1}^{\infty}\frac{1}{(\beta n-x_4)^5}.$$
Therefore, the integration over $d|{\bf x}|$ and the summation over winding modes yield 
$$\int_0^\infty dx x^2\sum\limits_{n=-\infty}^{+\infty}\frac{1}{[x^2+(\beta n+x_4)^2]^4}=
\frac{\pi}{32}\left\{\frac{1}{x_4^5}+\sum\limits_{n=1}^{\infty}\left[\frac{1}{(\beta n+x_4)^5}+
\frac{1}{(\beta n-x_4)^5}\right]\right\}.\eqno(A.2)$$
The coincidence of curly brackets in Eqs.~(A.1) and~(A.2) proves that the {\it ansatz} $\rho^{\rm pert}=N\omega^4$ 
captures the contribution of $\left<T_{12}(0)T_{12}(x)\right>_{\rm pert}$ correctly, with the corresponding normalization  factor being $N(T)=\pi^2A(T)(N_c^2-1)g^4(T)/192$. 

In this way, one isolates in the full Kubo formula simultaneously the perturbative 
$\omega^4$-part of $\rho^{\rm full}$ and the perturbative contribution to 
$\left<T_{12}(0)T_{12}(x)\right>_{\rm full}$. The remaining nonperturbative parts of 
$\rho^{\rm full}$ and $\left<T_{12}(0)T_{12}(x)\right>_{\rm full}$ are related to each other 
by means of Eq.~(\ref{1}).

\end{document}